# A FIBER-COUPLED DIAMOND QUANTUM NANOPHOTONIC INTERFACE

*Michael J. Burek[a], Charles Meuwly[a,b], Ruffin E. Evans[c], Mihir K. Bhaskar[c], Alp Sipahigil[c], Srujan Meesala[a], Denis D. Sukachev[c,d,e], Christian T. Nguyen[c], Jose L. Pacheco[e], Edward Bielejec[e], Mikhail D. Lukin[c], and Marko Lončar[a,†]*

[a.] John A. Paulson School of Engineering and Applied Sciences, Harvard University, 29 Oxford Street, Cambridge, MA 02138, USA

[b.] École Polytechnique Fédérale de Lausanne (EPFL), CH-1015 Lausanne, Switzerland

[c.] Department of Physics, Harvard University, 17 Oxford Street, Cambridge, MA 02138, USA

[d.] P. N. Lebedev Physical Institute of the RAS, Moscow 119991, Russia

[e.] Russian Quantum Center, Skolkovo, Moscow 143025, Russia

[f.] Sandia National Laboratories, Albuquerque, NM 87185, USA

[†] Corresponding author contact: E-mail: loncar@seas.harvard.edu. Tel: (617) 495-579. Fax: (617) 496-6404.

**ABSTRACT** – Color centers in diamond provide a promising platform for quantum optics in the solid state, with coherent optical transitions and long-lived electron and nuclear spins. Building upon recent demonstrations of nanophotonic waveguides and optical cavities in single-crystal diamond, we now demonstrate on-chip diamond nanophotonics with a high efficiency fiber-optical interface, achieving > 90% power coupling at visible wavelengths. We use this approach to demonstrate a bright source of narrowband single photons, based on a silicon-vacancy color center embedded within a waveguide-coupled diamond photonic crystal cavity. Our fiber-coupled diamond quantum nanophotonic interface results in a high flux of coherent single photons into a single mode fiber, enabling new possibilities for realizing quantum networks that interface multiple emitters, both on-chip and separated by long distances.

# 1. INTRODUCTION

Luminescent point defects ("color centers") in diamond provide a solid-state platform for the realization of scalable quantum technologies [1]. For instance, demonstrations that leverage the nitrogen vacancy (NV) center in diamond as a spin-photon interface [2-7], have included the entanglement of two distant solid-state qubits [8] and long-distance quantum teleportation [9]. While such advances have enabled recent tests of fundamental laws of nature [10], entanglement generation rates in these experiments are currently limited by the rate of coherent photon generation and collection. To develop a scalable architecture for the realization of quantum networks [11, 12], it will ultimately be necessary to engineer efficient single-photon emission into well-defined spatio-temporal modes [13].

Towards this goal, parallel efforts in the field of diamond nanofabrication and nanophotonics, have demonstrated on-chip low loss (~ db/cm) diamond waveguides and a wide range of high quality (Q) factor optical cavities [12, 14-22]. Recently, angled-etching nanofabrication [23, 24] has emerged as a scalable method for realizing nanophotonic devices from bulk single-crystal diamond substrates. Using this approach, we have demonstrated high Q-factor ($> 10^5$) diamond photonic crystal cavities (PCCs) [25] operating over a wide wavelength range (visible to telecom). Monolithic diamond PCCs fabricated by angled-etching are especially attractive for their compatibility with post-fabrication processing techniques necessary to stabilize implantation-defined color centers, i.e. high temperature annealing and acid treatments [26, 27]. Together, recent efforts in quantum science and nanoscale engineering of diamond have resulted in the demonstration of a solid-state single-photon switch based on a single silicon-vacancy (SiV) color center embedded in a diamond PCC, as well as observation of entanglement between two SiVs implanted in a single diamond waveguide [12]. As diamond nanophotonics continues to enable advances in other disciplines (including non-linear optics [28, 29] and optomechanics [30, 31]), the demand for scalable technology necessitates moving beyond isolated devices, to fully integrated on-chip nanophotonic networks in which waveguides route photons between optical cavities. Moreover, for applications involving single photons, such as quantum nonlinear optics with diamond color centers [12, 22], efficient off-chip optical coupling schemes are necessary to provide seamless transition of on-chip photons into commercial single mode optical fibers [32-34].

Herein, we demonstrate on-chip diamond nanophotonics integrated with a high efficiency fiber-optical interface, achieving greater than 90% power coupling at visible wavelengths. Our diamond nanophotonic structures utilize free-standing angled-etched waveguides (Figure 1), which retain low optical loss despite being physically supported through attachment to the bulk substrate [35], and are able to efficiently route photons on-chip [25]. The fiber-optical coupling scheme utilizes a single mode optical fiber, with one end chemically etched into a conical taper, to adiabatically transition guided light between on-chip diamond waveguides and off-chip optical fiber networks. Finally, we use our fiber-coupled diamond nanophotonics platform to demonstrate a bright source of narrowband single photons (near Fourier-limited at < 1 GHz bandwidth), based on a SiV center embedded within a waveguide-coupled diamond PCC. We achieve a cavity enhanced, coherent single photon generation rate of ~ 38 kHz, representing nearly an order of magnitude improvement compared to previous demonstrations of coherent zero-phonon-line (ZPL) photons collected from a single color center in diamond [5, 6, 36], with the additional advantage of providing these photons in a single-mode optical fiber. Our work will ultimately enable new possibilities for realizing quantum networks that interface multiple emitters, both on-chip and separated by long distances.

## 2. ON-CHIP DIAMOND NANOPHOTONIC STRUCTURES

Figure 1 displays a series of SEM images revealing diamond nanophotonic structures realized by angled-etching nanofabrication [23, 24] (see Supplementary Material for more details [37]). The nanophotonic systems consist of four key components: (1) freestanding diamond waveguides, (2) waveguide-coupled diamond nanobeam photonic crystal cavities (PCCs, Figure 1 (b)), (3) vertical waveguide support structures (Figure 1 (c)), and lastly, (4) freestanding diamond waveguide tapers (DWTs, Figure 1 (d) and (e)).

*2.1 Diamond nanobeam photonic crystal cavities*

PCCs enable high optical Q-factors while retaining wavelength-scale mode volumes [38], a key ingredient enabling strong light-matter interactions [11, 13]. Additionally, diamond nanobeam PCCs (see Figure 1 (b) inset and Ref. [37] for additional details) are readily integrated with on-chip waveguides, as the cavity architecture is naturally built into a suspended waveguide segment. This makes it possible to engineer waveguide-damped cavities where the cavity decay to the waveguide is much larger than scattering and absorption losses [39, 40]. We target the diamond PCC design used in this work to support a fundamental cavity mode resonance near $\lambda \sim 737$ nm, corresponding to the optical transition of the SiV center [41, 42]. Additionally, we focus on transverse-electric (TE) polarized cavity modes, which have their major electric field component mostly perpendicular to the $y = 0$ plane (see Figure 1 (b) inset for coordinate convention). From finite difference time domain (FDTD) electromagnetic simulations, the cavity mode volume was $V_{mode} \sim 0.52 \, (\lambda/n)^3$, where $n$ is the diamond refractive index, a four-fold improvement compared with previous results [12]. For fabricated structures, the number of Bragg mirror segments included in the nanobeam cavity was set to fix the total Q-factor of the device (as confirmed by FDTD simulations) to $Q_{total} \sim 10^4$, with cavity losses dominated by coupling to the feeding waveguide (intrinsic radiative losses were estimated to be $Q_{rad} \sim 2 \times 10^5$ by FDTD simulations). Additionally, to limit insertion (scattering) losses into the cavity, several holes that reduce quadratically in pitch and radii were appended to the ends of the PCC to ensure an adiabatic transition between the waveguide mode and Bloch mode of the Bragg mirror [43].

*2.2 VERTICAL WAVEGUIDE SUPPORT STRUCTURES*

Freestanding waveguides and PCCs realized by angled-etching ultimately require physical support through attachment to the bulk diamond substrate. To ensure robust mechanical performance while minimizing optical transmission loss, we employ vertical waveguide support structures [35], created by increasing the waveguide width by approximately 30 % of the nominal value and gradually tapering over 10 μm long straight portions (Figure 1 (c)). With angled-etching nanofabrication, wider waveguide sections require longer etch times to fully release from the bulk substrate. Consequently, wider waveguide sections remain attached to the substrate, resulting in a pedestal-like cross-section [25, 35]. At the same time, the adiabatically tapered support along the waveguide path minimizes optical transmission losses through the structure. We estimate the thickness of the diamond material supporting the widest portion of the waveguide (mid-support) to be < 100 nm, with prior measurements of transmission losses through such fabricated structures to be on order $\sim 10^{-3}$ dB per support [25].

*2.3 DIAMOND WAVEGUIDE TAPERS*

To efficiency couple light from our diamond nanophotonic structures, we adapt a fiber-optical interface previously developed for suspended silicon nitride nanophotonic systems [32]. Specifically, we employ a single-ended conical optical fiber taper (OFT, described in the following section) to make physical contact between the OFT tip and the on-chip freestanding DWT (Figure 1 (e)). Freestanding DWTs gradually change the effective refractive index of the waveguide mode along the propagation direction, such that nearly all optical power remains in the target eigenmode. To minimize coupling to higher-order modes, we exploit gradual tapering which fulfills the adiabatic condition [32]: $dn_{wvg}/dx < (2\pi/\lambda)|n_{eff,1}(x) - n_{eff,2}(x)|^2$, where $n_{wvg}$ is the effective index of the diamond waveguide, $n_{eff,i}$ is the effective index of the $i^{th}$ waveguide mode along the taper, and $\lambda$ = 737 nm is the free-space wavelength. Freestanding DWTs were designed to evolve from the nominal diamond waveguide width (~ 500 nm) down to a sub < 50 nm point, over a 20 μm length, yielding a final taper angle of ~ 2°. Importantly, the DWTs scale in all three dimensions as a result angled-etching nanofabrication, since the waveguide width defines its thickness via a constant etch angle [24].

## 3. EFFICIENT OPTICAL FIBER TAPER COUPLING

*3.1 SIMULATION OF COUPLING EFFICIENCY*

In the region of physical contact between the DWT and OFT (schematically represented in Figure 2 (a)), propagating guided modes couple via their evanescent fields, forming a hybridized "supermode" [32]. Figure 2(b) displays the calculated effective indices of a DWT physically coupled to a single-ended OFT (insets display cross-sectional eigenmode profiles obtained from simulation). With this geometry, the latter has $n_{eff}$ greater than ~ 1.28 over the entire length of the coupler, indicating that the optical mode remains well confined throughout the DWT-OFT structure. To confirm adiabatic mode transfer, we employ FDTD simulations to launch a propagating fundamental TE polarized mode down the diamond waveguide (Figure 2(c)), and monitor the power output in the $HE_{11}$ optical fiber mode after the coupling region. A power transfer of ~ 98 % is achieved for a 20 μm contact region (equal to the DWT length), corresponding to a coupling loss of less than 0.09 dB per facet. High efficiency coupling is also observed as the OFT overlap with the DWT is increased up to nearly 50 μm (Figure 2 (c)).

*3.2 OPTICAL FIBER TAPER FABRICATION*

Single-ended conical OFTs were fabricated by a wet etching technique, where commercial near-infrared single-mode optical fibers (Thorlabs S630-HP) were submerged in hydrofluoric acid (HF) to form the taper profile [44], as depicted in Figure 2(d). A layer of o-xylene was added on top of the HF to promote gradual taper formation via an oil/water interface meniscus that wicks up the length of the fiber. As the etch progresses and the fiber diameter shrinks, the height of the oil/water interface meniscus naturally decreases, resulting in a tapered fiber diameter over a length defined by the etch rate and initial fiber diameter. When the acid etches completely through the fiber diameter, the OFT self-terminates. Drawing the fiber out of the HF solution at a fixed rate further extends the taper length, enabling full control over the final taper angle. Moreover, this etching protocol readily extends to the simultaneous fabrication of many nominally identical OFTs. For OFT angles less than ~ 4°, the $HE_{11}$ fiber mode adiabatically transitions over the entire length of the OFT (on order ~ 10 mm) [32]. Figure 2(d) shows a SEM image of a representative OFT tip fabricated by HF etching, with a final taper angle of ~ 1.5°. Here, the significant surface roughness of the etched OFT (Figure 2 (e)) is a result of initial surface topography of the original commercial fiber [37]. However, the total distance over which the optical mode evanescently leaks outside of the OFT tip is less than ~ 100 μm. Thus, scattering losses due to fiber roughness were not detrimental to the final coupling efficiency.

*3.3 OPTICAL CHARACTERIZATION OF DIAMOND NANOPHOTONIC STRUCTURES*

The generalized optical fiber network used to collect reflection spectra from the waveguide-coupled diamond PCCs is shown in Figure 3 (a). Motorized stages precisely controlled the position of the single-ended OFT to bring it into physical contact with the DWT. A white-light laser source (NKT Photonics EXW-12) was coupled into the fiber network and launched into a 2x2 90:10 fiber coupler, with 90% of reflected light returned to the detection path. The transmitted source light polarization was adjusted with an inline fiber polarizer to ensure preferential coupling to the fundamental TE-polarized diamond waveguide mode and TE-polarized diamond PCC cavity resonances, and an optical spectrum analyzer (OSA) recorded reflection spectra. In addition, for a qualitative estimate of coupling efficiency over a broad wavelength range, the diamond PCC under test was replaced with a commercial fiber-coupled retroreflector (FR-63, Silicon Lightwave Technology, Inc.).

Figure 3(b) displays an optical micrograph of the on-chip diamond nanophotonic network coupled to the OFT tip with the source white light laser turned on. Light readily couples through the optical fiber taper tip to the on-chip diamond waveguide, with the diamond PCC acting as a mirror for almost the entire visible emission band. Light passed by the diamond photonic crystal cavity is scattered at the end of the diamond waveguide by an intentionally placed notch in the waveguide. A broadband reflection spectrum collected from a representative diamond PCC (Figure 3 (c), blue curve), exhibits a series of reflection dips attributed to localized TE-polarized cavity resonances, and demonstrates the broadband nature (over 200 nm) of this OFT-DWT coupling approach. A high-resolution spectrum of the fundamental diamond PCC mode at 733.9 nm (indicated by the green circle) is shown in Figure 3 (d), with a Lorentzian fit to the spectra yielding a Q-factor of $\sim 1.1 \times 10^4$. The on-resonance normalized reflection is $\sim 46\%$, which give an estimate of the Q-factor due to radiative losses or absorption of $\sim 3.4 \times 10^4$ [40]. Higher order longitudinal modes of the cavity are strongly waveguide damped, the second order diamond PCC resonance a $\lambda \sim 780$ nm has $Q \sim 1.2 \times 10^3$).

Away from the localized cavity resonances, we assume the diamond PCC ideally reflects all the TE-polarized light coupled to the diamond waveguide, allowing for an accurate calculation of the coupling efficiency. The off-resonant input power ($P_{in}$) and the reflected powers after the 2x2 coupler ($P_r$) are measured with a calibrated power meter. The normalized reflection is thus given by $P_r/P_{in} = \eta_c^2 \eta_{BS} \eta_m \eta_{FC}$, where $\eta_c$ is the OFT-DWT coupling efficiency, $\eta_{BS}$ is the calibrated coupling ratio of the 2x2 fiber coupler, $\eta_m$ is the reflection of the nanobeam cavity Bragg mirror (assumed to be ~ 1), and $\eta_{FC}$ ~ 92 % is the coupling efficiency of the FC-FC fiber coupler. Neglecting other losses in the diamond waveguide, we estimate a coupling efficiency $\eta_c$ ~ 91 % (measured with a pigtailed laser diode operating at wavelength of ~ 705 nm). We carried out a series of similar measurements [37] on waveguide-coupled diamond PCCs fabricated for operation at telecom wavelengths (~ 1480 to 1680 nm). Here, a maximum measured coupling efficiency of ~ 96 %, demonstrates the broadband nature of both diamond nanophotonics and this adiabatic OFT-DWT coupling scheme.

## 4. EFFICIENT GENERATION AND COLLECTION OF NARROWBAND SINGLE PHOTONS

We utilize our described diamond nanophotonic structures to implement a bright source of narrowband single photons suitable for use in quantum information protocols. SiV color centers are embedded at the center of waveguide-coupled diamond PCCs via focused ion beam (FIB) implantation [12, 45, 46]. SiV centers are excited from free space using a scanning confocal microscope [12]. Fluorescence is detected in the waveguide using the fiber-coupled interface (depicted in Figure 4(a)). In order to isolate narrow optical transitions, we cool the device to a temperature of ~ 5 K within a liquid helium continuous flow cryostat, at which an optical Λ-system formed by the SiV spin-orbit eigenstates is accessible (Figure 4 (b), left inset) [41, 42]. We use resonant excitation on the $|u\rangle$ to $|e\rangle$ branch to generate Raman fluorescence on the $|e\rangle$ to $|c\rangle$ transition. One benefit of this scheme is that, by tuning the frequency of the driving laser, the frequency of the emitted photons can be tuned by more than 10 GHz [12]. We selectively collect this spectral component of the fluorescence using a home-built Fabry-Perot (FP) cavity (finesse ~ 100) as a filter. By scanning the resonance frequency of the FP-cavity across the $|e\rangle$ to $|c\rangle$ transition, we measure an upper bound on the linewidth of collected photons of $1.33 \pm 0.02$ GHz, set by the FP cavity (Figure 4(b)). With the FP cavity fixed on resonance with the $|e\rangle$ to $|c\rangle$ transition, we measure the saturation of Raman fluorescence (Figure 4(c)) when the PCC is tuned onto resonance with the $|e\rangle$ to $|c\rangle$ transition. Here $\kappa$, the PCC linewidth measured via a tunable laser is approximately 39 GHz (Q ~ $10^4$). Tuning of the cavity resonance is achieved via controlled condensation of an inert gas onto the device [12]. We measure a maximum Raman photon detection rate of ~ 38 kHz of when the PCC is on resonance. The fluorescence consists of single photons, as confirmed by anti-bunching in a fluorescence autocorrelation measurement (Figure 4(c), right inset). Our cavity enhanced, coherent photon generation rate represents a significant improvement over the state of the art, when compared against previous demonstrations of coherent ZPL photons collected from a single diamond color center. The spectral purity (< 1 GHz bandwidth) of our collected photons is an essential element of schemes requiring indistinguishable photons, including entanglement generation (the rate of which is typically proportional to the square of the photon generation rate [8]). Our work represents a significant improvement over the state of the art for narrow-band photon generation using color centers in diamond, where collection rates for spectrally pure ZPL photons from an individual SiV center [36] or NV center [5, 6] of only ~ 6 kHz and 0.2 – 1.1 kHz were observed, respectively.

Our photon detection rate is limited primarily by finite transmission through the FP cavity filter and related free-space optics (~ 11% transmission) after collection into the OFT. Accounting for these losses, we estimate a lower bound on the collection rate of unfiltered single photons to be ~ 0.45MHz in the single-mode optical fiber. Additionally, in the case of the particular waveguide-coupled diamond PCC device measured, the OFT-DWT coupling efficiency was limited to approximately 22 to 25%, primarily due to mechanical failure of the specific DWTs used in this experiment, and the limited degrees of freedom available for accurate fiber positioning in our cryogenic apparatus. While this OFT-DWT coupling efficiency is less than what we have observed in ambient conditions, typical collection efficiencies into a single-mode fiber are < 10 % even in the best-case scenario of a high-NA objective and a solid immersion lens (SIL) machined into the diamond substrate [9], and typically < 1 %. Mitigating these current issues with improved post-fabrication sample processing and cryogenic OFT-coupling techniques, we expected to increase our detection rate of spectrally filtered ZPL photons beyond the ~ 100 kHz limit, assuming similar transmission losses through the FP cavity filter. Finally, in this continuous-driving Raman scheme, the ultimate limit to the single-photon emission rate is determined by the phonon-limited relaxation time (~ 40 ns) [47] between the ground states, which is required to reset population into $|u\rangle$ after emission of a Raman photon. In the future, this may be overcome in a pulsed excitation scheme in which the SiV is optically initialized in the state $|u\rangle$ before the Raman excitation pulse is applied.

## 5. CONCLUSIONS

In summary, we demonstrate on-chip diamond nanophotonic structures with a high efficiency fiber-optical interface, achieving > 90% power coupling at visible wavelengths. Our diamond nanophotonic networks utilize freestanding angled-etched waveguides, which retain low optical loss despite being physically supported through attachment to the bulk substrate, and are able to efficiently route photons on-chip. The fiber-optical coupling utilizes a single mode optical fiber, with one end chemically etched into a conical taper, to adiabatically transition guided light between on-chip diamond waveguides and off-chip optical fiber networks. With a SiV center embedded within our waveguide-coupled diamond PCC, we demonstrate a ~ 38 kHz flux of spectrally narrow single photons (< 1 GHz bandwidth), efficiently coupled to single-mode optical fiber. Our bright and narrowband fiber-integrated diamond nanophotonic quantum node is of immediate technological significance to applications in quantum optics [12]. Combined with advances in quantum control of the diamond SiV center [48-50] and schemes for improved spin coherence times [47], this platform opens up new possibilities for realizing large-scale systems involving multiple emitters strongly interacting via photons [11, 13].

## ACKNOWLEDGEMENTS

The authors acknowledge V. Venkataraman for useful discussion in the course of preparing this report and D. Perry for performing the focused ion beam implantation. This work was supported in part by the AFOSR Quantum Memories MURI (grant FA9550-12-1-0025), ONR MURI on Quantum Optomechanics (Grant No. N00014-15-1-2761), NSF QOP (grant PHY-0969816), NSF CUA (grant PHY-1125846), and the STC Center for Integrated Quantum Materials (NSF Grant No. DMR-1231319). Ion implantation was performed with support from the Laboratory Directed Research and Development Program and the Center for Integrated Nanotechnologies at Sandia National Laboratories, an Office of Science facility operated for the DOE (contract DE-AC04-94AL85000) by Sandia Corporation, a Lockheed Martin subsidiary. Device fabrication was performed in part at the Center for Nanoscale Systems (CNS), a member of the National Nanotechnology Infrastructure Network (NNIN), which is supported by the National Science Foundation under NSF award no. ECS-0335765. CNS is part of Harvard University.

**FIGURES**

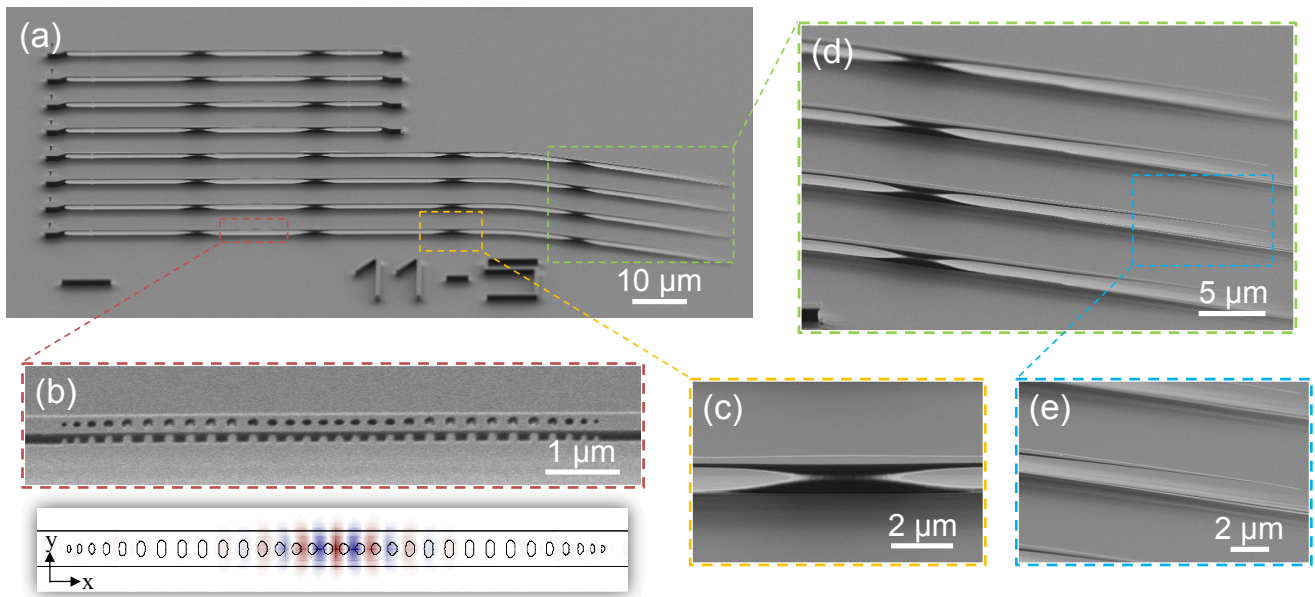

**Figure 1 | On-chip diamond nanophotonic structures.** SEM images of (a) an array of diamond nanophotonic structures, fabricated using angled-etching techniques [37]. The four key components in each device are: (1) freestanding diamond waveguides, (2) integrated diamond nanobeam photonic crystal cavities (panels (b) with the top down electric field profile of the fundamental optical cavity mode inset), (3) vertical waveguide support structures (panel (c)), and lastly, (4) freestanding diamond waveguide tapers (panels (d) and (e)).

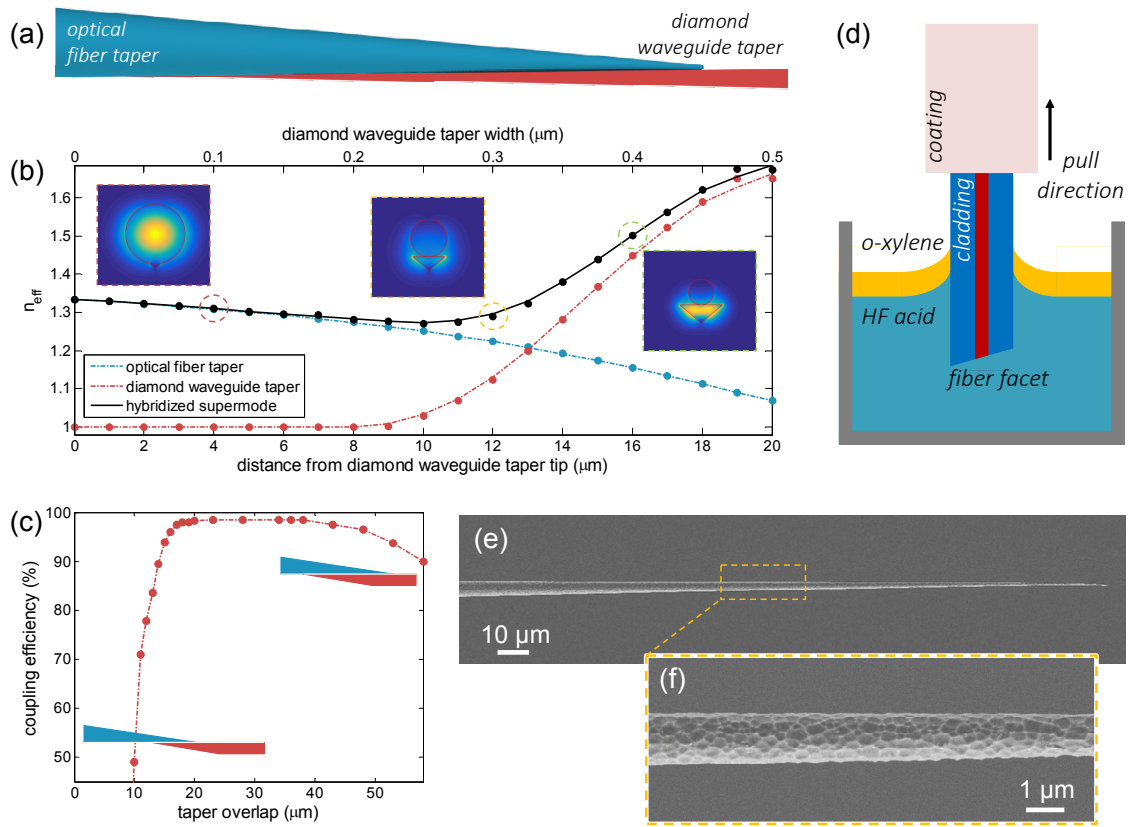

**Figure 2 | Single-mode optical fiber taper fabrication.** (a) Schematic of fiber-waveguide taper adiabatic coupling. The single-ended conical optical fiber taper (OFT, blue) is physically contacted with a diamond waveguide taper (DWT, red). (b) Effective indices ($n_{eff}$, calculated via an eigenmode solver) of the OFT and transverse-electric (TE) polarized DWT modes for taper angle of 1.5° in both cases. Inset: cross-sectional energy density profile ($|E|^2$) obtained from eigenmode simulations at the indicated points along the physically overlapped tapers. (c) Simulated coupling efficiencies for transmission from the fundamental diamond TE-polarized waveguide mode to the optical fiber $HE_{11}$ mode as a function of fiber-waveguide taper overlap for a 20 μm long DWT. (d) Schematic of fabrication of single-ended conical OFTs by hydrofluoric acid etching. (e) SEM image of a fabricated OFT tip, with (f) a zoomed in image of the chemically etched surface roughness.

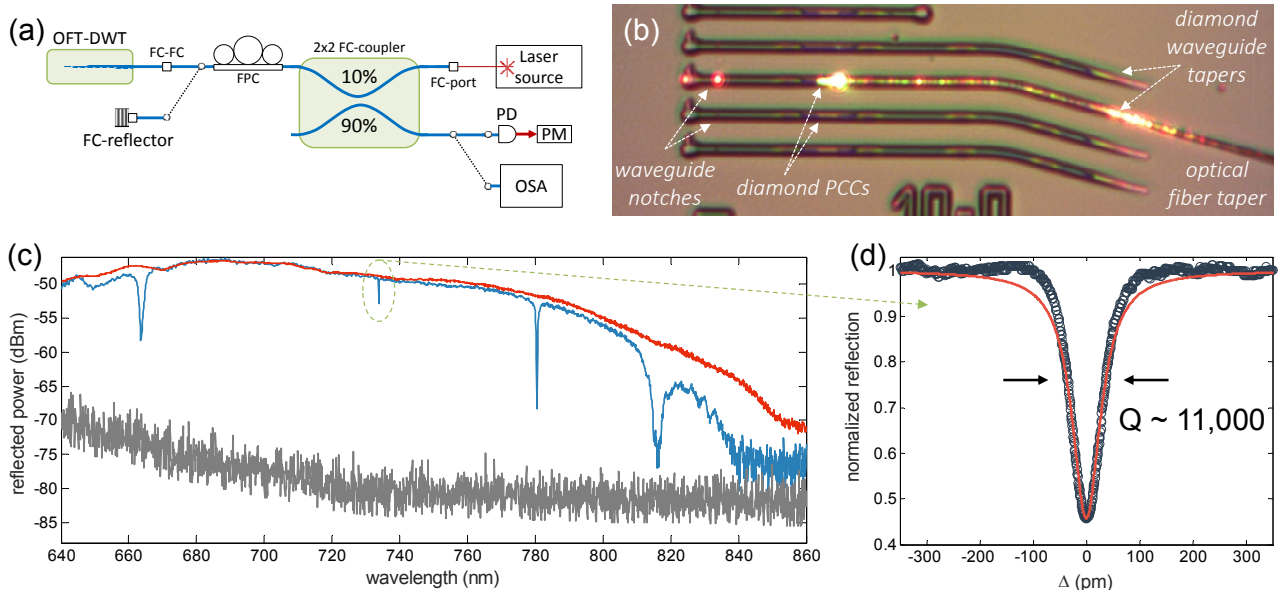

**Figure 3 | Optical characterization of diamond nanophotonic structures.** (a) Schematic of the visible band fiber optical characterization set up. (b) Optical micrograph of a single-ended optical fiber taper (OFT) in contact with a diamond waveguide taper (DWT) under white light illumination. (c) Broadband reflection spectra of diamond photonic crystal cavity (PCC, blue curve) and a commercial fiber-coupled retroreflector (red curve). The noise floor of the optical spectrum analyzer (OSA) is shown in grey. (d) Normalized high resolution spectrum of the fundamental diamond nanobeam PCC mode centered at $\lambda \sim 733.9$ nm. A Lorentzian fit (red curve) yields a Q-factor estimate of $Q \sim 1.1 \times 10^4$.

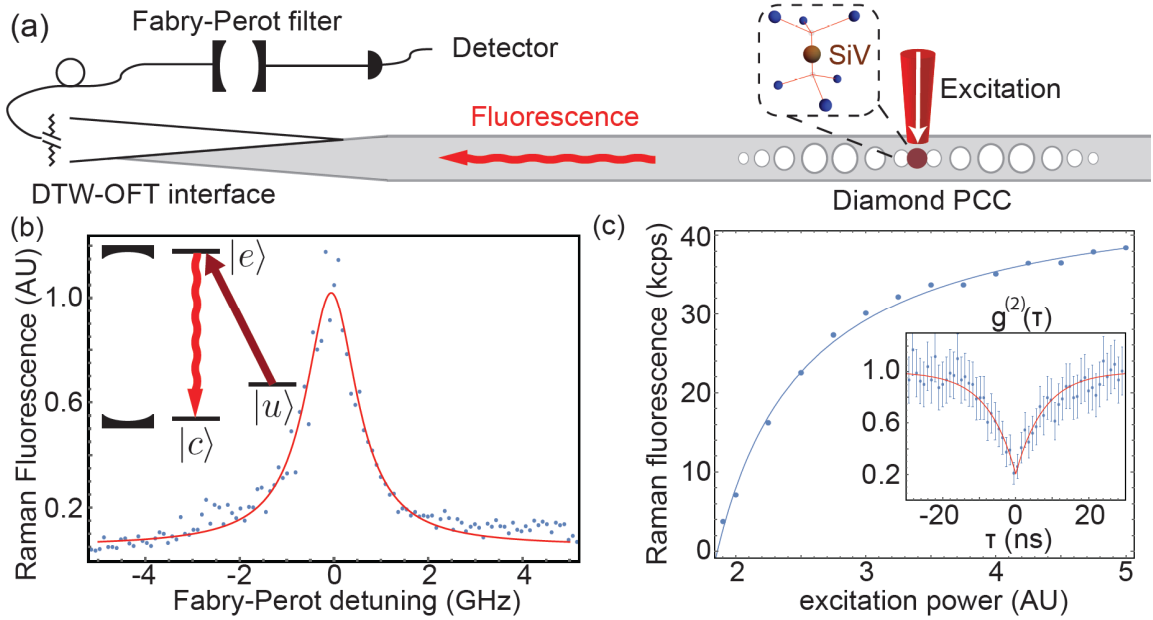

**Figure 4 | Efficient generation and collection of narrowband single photons.** (a) A diamond SiV color center embedded inside a diamond photonic crystal cavity (PCC) is excited from free space, and fluorescence in the diamond waveguide is collected into the optical fiber taper. A Fabry-Perot (FP) cavity is used to measure the spectrum of the emitted photons. (b) Spectrum of collected photons. The excitation laser is resonant with the $|u\rangle$ to $|e\rangle$ transition. The transmitted Raman fluorescence is recorded as the resonance frequency of the FP cavity is scanned across the $|e\rangle$ to $|c\rangle$ transition to obtain an upper bound on the linewidth of detected photons of 1.33 +/- 0.02 GHz. (c) Saturation measurement of Raman photons with the PCC resonant with the $|e\rangle$ to $|c\rangle$ transition. We subtract the linear background set by the excitation laser, detecting a narrow-band single photon flux of at least 38 kHz when the PCC is on resonance. Inset: anti-bunching of $g^{(2)}(0) = 0.21$ +/- 0.09 in the Raman fluorescence autocorrelation confirms the generation of single photons.

**SUPPLEMENTARY MATERIAL**

**i) Angled-etching nanofabrication of diamond nanophotonic structures**

Our angled-etching approach, schematically depicted in Figure S1 (a), with corresponding SEM images displayed in Figure S2 subpanels (b) to (e), employs anisotropic oxygen-based plasma etching at an oblique angle to the substrate surface, resulting in suspended structures with triangular cross-section [1, 2]. Angled-etching is performed in a standard inductively coupled plasma-reactive ion etcher (ICP-RIE, UNAXIS Shuttleline). However, to modify the trajectory of the incident plasma ions towards the sample surface, the diamond substrate is housed within a specifically designed aluminum Faraday cage. Details regarding the Faraday cage construction are found elsewhere [3]. Note, a ~ 50° etch angle (designated as the half angle at the bottom apex of the triangular cross-section, see inset Figure S1 (a)) was used to fabricate freestanding diamond waveguides and nanobeam photonic crystal cavities (PCCs).

Single-crystal diamond substrates (CVD grown, <100>-normal oriented, < 5 ppb [N], Element Six), were first polished (performed commercially by Delaware Diamond Knives) to a surface roughness < 5 nm RMS. After polishing, substrates were cleaned in a boiling mixture consisting of equal parts sulfuric acid, nitric acid, and perchloric acid (referred to as a "tri-acid clean" hereafter). Prior to device fabrication, ~ 6 microns of the diamond substrate surface was removed to eliminate any polishing-induced mechanical strain. This surface preparation step was performed via a ICP-RIE plasma process, and included a 30 minute etch with the following parameters: 400 W ICP power, 250 RF power, 25 sccm Ar flow rate, 40 sccm $Cl_2$ flow rate, and 8 mTorr chamber pressure, followed by a second 30 minute etch with the following parameters: 700 W ICP power, 100 RF power, 50 sccm O2 flow rate, and 10 mTorr chamber pressure. The substrate surface roughness following this step was reduced to < 1 nm RMS (confirmed by AFM [4]).

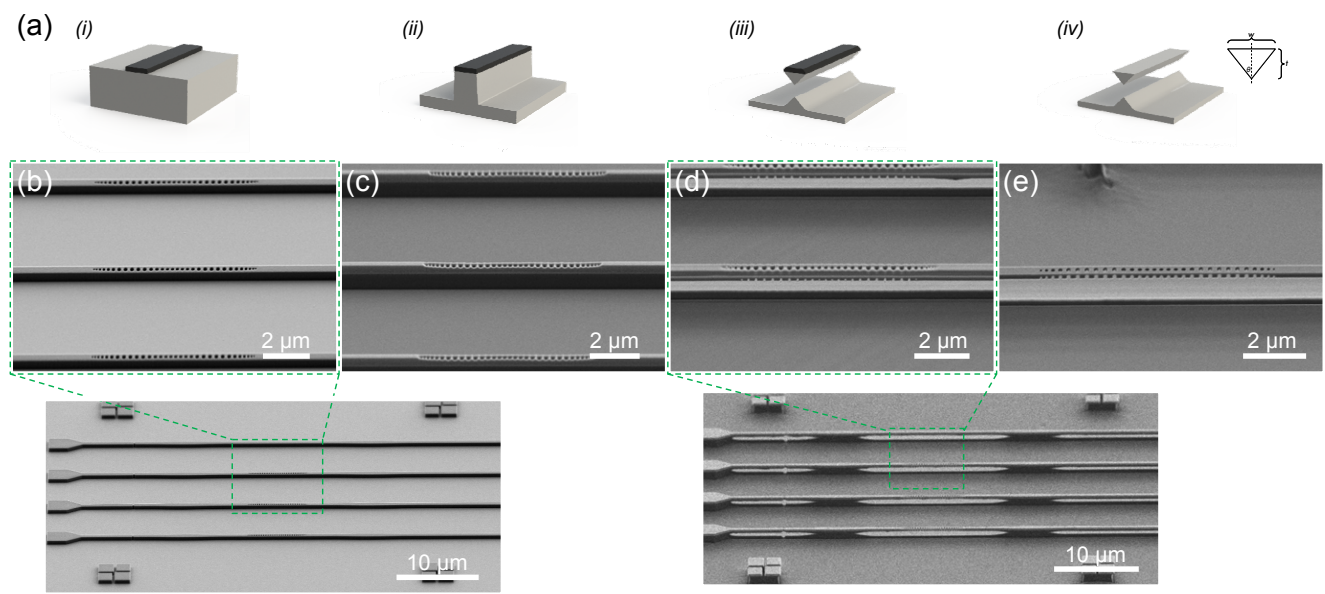

**Figure S1 | Angled-etching nanofabrication of diamond nanophotonic structures.** (a) Illustration of the angled-etching fabrication scheme, with corresponding SEM images: (i) define an etch mask on substrate via standard fabrication techniques (panel (b)), (ii) transfer etch mask pattern into the substrate by conventional top down plasma etching (panel (c)), (iii) employ angled-etching to realize suspended nanobeam structures (panel (d)), (iv) remove residual etch mask (panel (e)).

Electron beam lithography and a negative tone hydrogen silsesquioxane (HSQ) based resist (FOX®-16 from Dow Corning) resist was used to create a silica etch mask on the prepared diamond substrates. Exposed HSQ was developed in tetramethylammonium hydroxide (TMAH, 25% diluted solution), and the final resist thickness was approximately ~ 650 nm. The prescribed silicon etch mask pattern was then transferred into the diamond substrate to a depth of ~ 600 nm, via a conventional top down anisotropic plasma etch (also performed in the UNAXIS Shuttleline ICP-RIE) with the following parameters: 700 W ICP power, 100 RF power, 50 sccm $O_2$ flow rate, 2 sccm Ar flow rate, and 10 mTorr chamber pressure. Following this, angled-etching was performed using the same ICP-RIE parameters as the initial top down etch, but with the sample housed within the aforementioned Faraday cage [3]. Remaining etch mask after completion of the oxygen-based plasma etching was removed in concentrated hydrofluoric acid, immediately followed by cleaning in piranha solution. The sample was transferred from acid into deionized water, then methanol, and finally blown dry in a stream of filtered nitrogen gas. A similar cleaning procedure was used after the diamond substrate was exposed to any wet processing steps (such as those described below). Critical point drying (CPD) was not used in the fabrication of samples with freestanding diamond waveguides reported throughout this work, resulting in a systematic reduction of diamond waveguide taper yield in repeated solution based cleanings. DWTs which fail mechanically (i.e. snap down and contact the underlying substrate) resulted in a reduced coupling efficiency to optical fiber tapers. Efforts to incorporate CPD as a routine step in the fabrication and processing of diamond nanophotonic networks with freestanding DWTs remain ongoing.

**ii) Focused ion beam implantation of silicon vacancy (SiV) centers into diamond nanophotonic structures**

With successful fabrication of diamond nanophotonic structures, SiV color centers were created in the approximate center of the diamond photonic crystal cavities (PCCs) by targeted implantation using a focused ion beam (FIB) [5-7]. Using Sandia National Laboratories' nanoImplanter (a custom FIB implantation system made by the A&D Corporation of Tokyo), the diamond PCCs were implanted with 100 keV $Si^+$ ions from an AuSiSb liquid metal alloy ion source. Each implantation site was targeted with between 10 to 500 ions using a 40 nm beam spot (equivalent to an ion fluence of $8\times10^{11}$ to $4\times10^{13}$ ions/cm$^2$). The ion fluence was controlled by a combination of beam current and dwell time. With the instrument's laser interferometry controlled stage and a Raith Elphy Plus pattern generator, the implantation position was aligned using patterned alignment markers to < 1 nm. This combination of controlled ion fluence and positioning allows for precise SiV formation within the diamond PCC. The expected error associated with the ion fluence is approximately the square root of the average number of implanted ions as the ions follow a Poisson distribution function. The expected positioning error is dominated by the ion spot size (40 nm) and the straggle of the implanted ions, predicted to be 68±16 nm in depth and ±13 nm laterally from SRIM simulations. The positioning error is approximately a factor of two smaller than the relevant cavity mode dimensions.

FIB implantation is followed by a three-stage UHV anneal (maximum pressure $5 \times 10^{-9}$ Torr) at 400°C (3°C per minute ramp, 8 hour dwell time), 800°C (1°C per minute ramp, 8 hour dwell time), and 1200°C (1°C per minute ramp, 4 hour dwell time). This annealing introduces a small amount of graphitic carbon on the surface of the sample. The tri-acid clean described in the previous section is repeated after annealing to remove this carbon. The effect of each post-implantation step is described in more detail in Ref. [8, 9]. The diamond PCC used in the experiment was implanted with 350 $Si^+$ ions and contains more than 5 SiV centers after high temperature annealing. These measurements demonstrate approximate 1.5% conversion yield from $Si^+$ ions to SiV centers, comparable to what has been realized in the bulk [8].

**iii) Diamond nanobeam photonic crystal cavity design**

The diamond nanobeam photonic crystal cavity (PCC) design used in this work consists of a triangular cross-section diamond waveguide perforated with a chirped lattice of elliptically-shaped air holes. The nominal unit cell (Figure S2 (a)), supports two polarizations of guided Bloch modes, as revealed in the corresponding photonic bandstructure (Figure S2 (b)) for a nominal unit cell with θ = 50° and ($a$, $w$, $d_z$, $d_x$) = (270, 470, 235, 120) nm). Here, transverse electric (TE, solid black lines) and transverse magnetic (TM, dashed blue lines) guided modes, exist below a continuum of leaky and radiation modes bounded the light line (grey shaded region), and give rise to symmetry based quasi-bandgaps sufficient to realize highly localized resonances. In our convention, the TE(TM)-polarized modes have the major electric field component mostly perpendicular (parallel) to the $y = 0$ plane (see Figure S2(a) for coordinate conventions). In other words, the electric field of a TE (TM) mode is mostly perpendicular (parallel) to the $z = 0$ plane. For our purposes, we focused on localized TE-polarized cavity modes near frequencies of $\omega_o/2\pi \sim 405$ THz ($\lambda \sim 740$ nm), and chose unit cell dimensions accordingly.

In our design, and the optical cavity is localized by quadratically reducing the hole-to-hole spacing ("periodicity") and air hole major radius ($d_y$) over the 5 air holes immediately adjacent to the $x = 0$ mirror plane of the lattice of air holes, as depicted in Figure S2(c). This gradual tapering raises the X-point frequency of the 1$^{st}$ order TE-polarized guided mode ('dielectric" band) into the Bragg mirror quasi-bandgap. This results in a localized defect mode with an attenuation profile which minimizes the spatial Fourier harmonics of the cavity mode inside the lightcone, thereby maximizing its radiation Q-factor ($Q_{rad}$). The total cavity loss is comprised of both radiation losses into free-space (denoted $Q_{rad}$) and coupling losses to the feeding waveguide (denoted $Q_{wg}$). Conveniently, $Q_{wg}$ may be independently increased by simply adding more air holes to the Bragg mirror along the waveguide. To calculate the cavity design optical resonances and losses, we employed finite-difference time-domain (FDTD, Lumerical Solutions Inc.) methods.

By the angled-etching process, the diamond nanobeam thickness ($t$) is intrinsically linked to its width ($w$) by the etch angle ($\theta$), via the relationship $t = w/(2 \tan \theta)$. Thus, with the scale invariance of Maxwell's equations, global scaling of the diamond PCC dimensions (with a fixed etch angle) results in tuning of the PCC resonance while maintaining all cavity figures of merit (i.e. Q-factor and mode volume). For this reason, we parameterized the diamond PCC design by the target fundamental TE-polarized cavity mode resonance wavelength, $\lambda_{TE}$. Our final design, as confirmed by FDTD simulations, has a width $w = 0.678\lambda_{TE}$, lattice constant and elliptical air hole diameters in the Bragg mirror of $(a, d_x, d_y) = (0.390\lambda_{TE}, 0.173\lambda_{TE}, 0.339\lambda_{TE})$, and lattice constant and elliptical air hole diameters at the center of the PCC of $(a, d_x, d_y) = (0.300\lambda_{TE}, 0.173\lambda_{TE}, 0.201\lambda_{TE})$. This quadratically tapering major diameter is schematically displayed in Figure S3 (c). For the unit cell dimensions used to calculate the bandstructure in Figure S2 (b), the fundamental TE-like cavity resonance is located at $\lambda_{TE} = 693$ nm, which is designated by the dashed red line in Figure S2 (b). Fabricated structures utilized this design scaled to target $\lambda_{TE} \sim 740$nm.

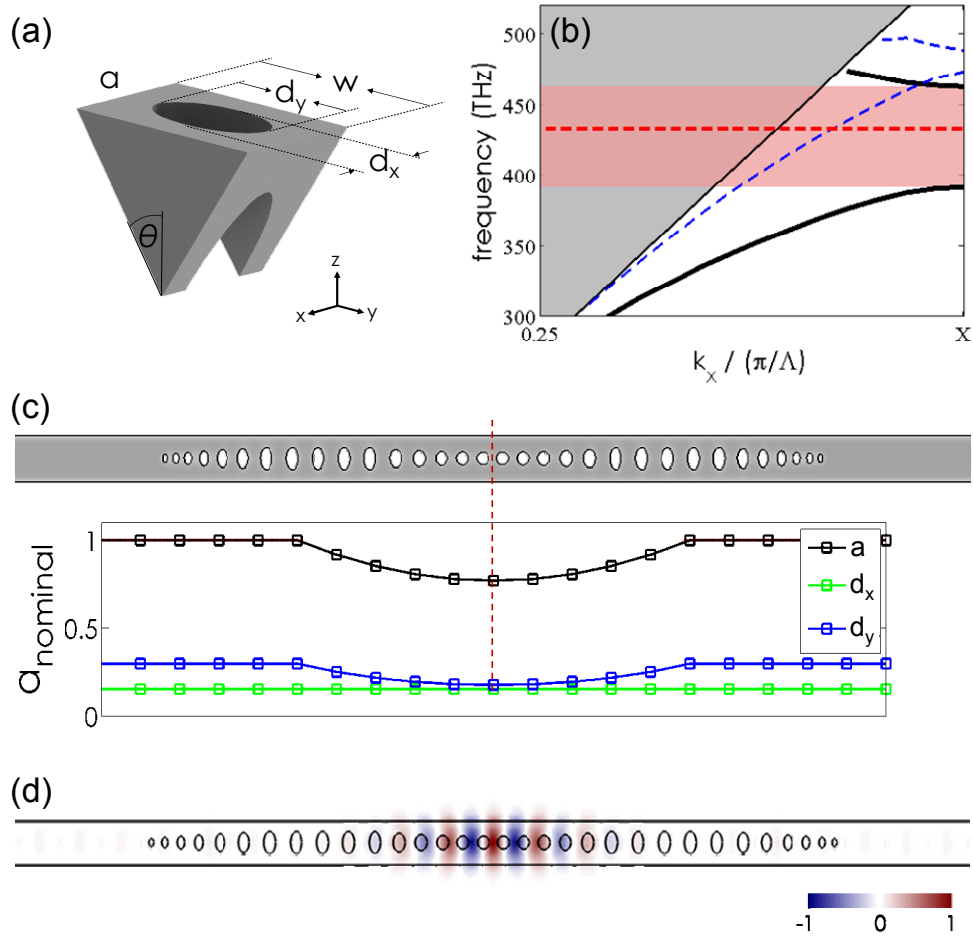

**Figure S2 | Diamond nanobeam photonic crystal cavity (PCC) design.** (a) Schematic representation of a nominal unit cell, parameterized by the etch angle ($\theta$), width (w), lattice constant (a), and major and minor elliptical air hole diameters ($d_x$, $d_y$). Corresponding (b) photonic bandstructure of a with $\theta = 50°$ and ($a$, $w$, $d_x$, $d_y$) = (270, 470, 120, 235) nm. In (b), the grey shaded region indicates the continuum of radiation and leaky modes that exist above the light line for the structure. Below the light line, supported transverse electric (TE) and transverse magnetic (TM) guided modes are indicted by solid black and dashed blue lines, respectively. A quasi-bandgap based on symmetry for the TE-polarized guided modes is indicated by the pink shaded region. (c) Schematic of the air hole array cavity design with the lattice constant and elliptical air hole diameters plotted as a function of Bragg mirror hole-to-hole spacing ($a_{nominal}$). (d) Normalized top plane optical electric-field profile of the fundamental TE-polarized diamond PCC mode for the design designated in (c). The fundamental cavity resonance at $\lambda_{TE} = 693$ nm is designated by the dashed red line in (b).

Also from FDTD simulations, the diamond PCC mode volume and radiation Q-factor (assuming many Bragg mirror segments) are: $V_{mode} \sim 0.52\ (\lambda/n)^3$ and $Q_{rad} \sim 2 \times 10^5$, respectively. Since radiation losses of the fabricated structure are typically limited by scattering due to fabrication imperfections (i.e., surface roughness), the number of mirror segments in the final cavity design was chosen to fix $Q_{wg}$ to be on same order as the $Q_{tot}$, thus achieving a waveguide damped cavity where the majority of optical energy leaks into the feeding diamond waveguide. This corresponded to approximately five air holes in the Bragg mirror segment of the diamond PCC, fixing $Q_{wg}$ at $\sim 10^4$. Additionally, to limit insertion (scattering) losses at the Bragg mirror/waveguide interfaces, several holes that reduce quadratically in pitch and radii were appended to the ends of the PCC in order to promote an adiabatic transition between the waveguide mode and Bloch mode of the Bragg mirror [10].

**iv) Optical characterization of diamond nanophotonic structures operating at telecom wavelengths**

In addition to optical characterization of diamond nanophotonic networks fabricated for operation at visible wavelengths (Figure 1 and Figure 3 of the main text), a series of identically designed devices were fabricated for operation over the telecom C-band and L-band (~ 1480 to 1680 nm). This was achieved by a simple linear scaling of all device design parameters (both optical cavity and waveguide taper dimensions) described in the previous section and in the main text. Such diamond nanophotonic systems operating at telecom wavelengths are relevant for applications in non-linear optics [11, 12] and optomechanics [13, 14]), and demonstrate the broadband nature of both the diamond material system and the optical fiber taper (OFT) adiabatic coupling scheme presented in this work.

To characterized fabricated on-chip diamond nanophotonic structures for operating at telecom wavelengths, two tunable lasers (Santec TSL-510, tuning range from 1480 to 1680 nm) were used, along with an inline fiber polarizer, and high gain InGaAs photodetector (EO Systems, IGA1.9-010-H) and power meter to record reflection spectra. Laser light was coupled to the diamond photonic crystal cavity (PCC) under test using a similar fiber optical characterization set up as displayed in Figure 3(a) of the main text. All collected spectra were normalized by reflected data collected from a fiber-coupled retroflector (Thorlabs, P5-SMF28ER-P01-1) with > 98% reflection over the laser tuning bandwidth.

Figure S3 (a) displays an optical micrograph of the coupling region where the OFT tip is in physical contact with the diamond waveguide taper (DWT), with an estimated taper overlap length of ~ 30 μm. A normalized reflection spectra collected from a representative waveguide-coupled diamond PCC is shown in Figure S3 (b). Here, the red curve corresponds to reflection spectrum associated with the transverse electric (TE) polarized diamond waveguide mode, while the blue curve corresponds to the transverse magnetic (TM) waveguide polarization. A set of reflection dips observed in the red curve are attributed to localized resonances of the diamond PCC with TE polarization. For the particular diamond PCC design (described in the previous section), a TM passband (high transmission, low reflection) exists in the spectral region associated with the TE stopband (high reflection). Broad resonances observed in the reflection spectrum associated with TM waveguide polarization are attributed to diamond PCC resonances of TM polarization [3] (not investigated in this work).

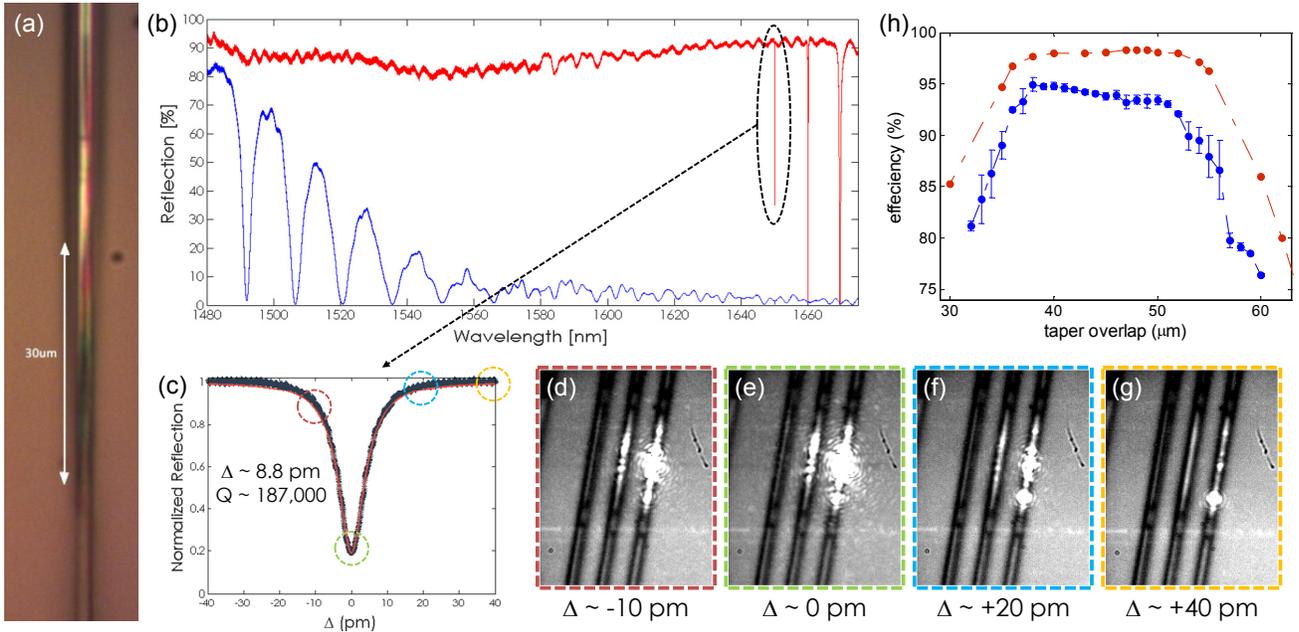

**Figure S3 | Optical characterization of diamond nanophotonic networks operating at telecom wavelengths.** (a) Optical micrograph of the fiber-waveguide coupling region. (b) Normalized broadband reflection spectra of a diamond nanobeam PCC operating at telecom wavelengths for both transverse electric (TE, red curve) and transverse magnetic (TM, blue curve) waveguide polarizations. Reflection dips correspond to cavity modes, with a high-resolution spectra of the fundamental TE-polarized cavity mode included in (c). Corresponding overhead infrared camera images of the diamond PCC (right most device) taken with the laser wavelength tuned (d) –10 pm, (e) 0 pm, (f) + 20 pm, and (g) + 40 pm from the cavity resonance. Light scattered out of the waveguide is visible when the laser is tuned close to the cavity resonance, confirming the spectral feature is a cavity mode. Note, that only an individual diamond PCC is pumped in this measurement, and illumination of the adjacent diamond PCC is due to in plane scattering of light. (h) Simulation and experimentally measured coupling efficiencies between the fundamental diamond TE waveguide mode to the optical fiber fundamental $HE_{11}$ mode as a function of OFT-DWT overlap. Note, here the DWT was approximately 40 microns.

In Figure S3 (d), a high-resolution scan of the fundamental TE-polarized cavity mode at 1638.1 nm is shown, with a Lorentzian fit to the spectra yielding a total Q-factor of ~ 187,000, with ~ 20% on-resonance reflection (nearly 80% extinction). Higher order longitudinal modes of the cavity are near-perfectly waveguide damped, yielding critically coupled resonances. Figure S3 (d) to (g) show optical micrographs collected overhead of the cavities with an infrared camera. In each image, the laser position is tuned relative to the fundamental TE cavity mode (laser detuning confirmed via a wavemeter), with light scattered by the cavity visible when the laser is tuned closed to resonance. With the laser tuned off-resonance, only a small amount of scattered light at the entrance to the cavity structure is visible, which is likely the result of small insertion losses between the propagating waveguide mode and the diamond PCC structure.

Away from the localized TE cavity resonances, we assume the diamond nanobeam cavity ideally reflects all the light, allowing for a calculation of the coupling efficiency. To do so, the laser is tuned just off-resonance from the second order TE cavity mode, and the input polarization is adjusted to maximize the reflected signal. Following the procedure described in the main text, the OFT-DWT coupling efficiency was extracted from the ratio of input laser power to reflected power, accounting for all calibrated losses. For the measurement displayed in Figure S3 (b), we estimate a coupling efficiency $\eta_c$ ~ 96 % at a laser wavelength near 1650 nm. Note that the broadband spectrum in Figure S3 (b) plots a reflected signal normalized by the broadband fiber retroreflector, and is thus a secondary measurement of coupling efficiency. Here, the normalized reflection (assuming near unit reflection of the fiber retroreflector) is ~ $P_r/P_{in} = \eta_c^2$, since the fiber retroreflector and OFT-DWT system experience the same system losses From the plot in Figure S3 (b), an estimated coupling efficiency of ~ 94 % is extracted at ~ 1650 nm, in excellent agreement with the measurement made with the power meter. The variable reflected signal in Figure S3 (b) is attributed to wavelength-dependent changes in fiber polarization.

To independently confirm robust high-efficiency coupling between the OFT and DWT, the coupling efficiency was extracted as a function of approximate OFT-DWT overlap, as plotted in Figure S3 (e). This measurement was performed with a DWT fabricated for operation at telecom waveguides (approximate length of ~ 40 microns), which allowed for a reasonably accurate estimate the OFT-DWT overlap length. Here, the error bars correspond to a standard deviation across four individual measurements. The trend in measured coupling efficiencies closely follows that expected from FDTD simulations, and demonstrated the robust nature of OFT-DWT alignment.

## SUPPLEMENTARY REFERENCES